\documentclass[aps,preprint]{revtex4}%
\usepackage{amsfonts}
\usepackage{amsmath}
\usepackage{amssymb}
\usepackage{graphicx}%
\begin{document}
\title{Influence of separating distance between atomic sensors for gravitational wave
detection }
\author{Biao Tang$^{a}$}
\author{Baocheng Zhang$^{a,b}$}
\email{zhangbc.zhang@yahoo.com}
\author{Lin Zhou$^{a}$}
\author{Jin Wang$^{a}$}
\author{Mingsheng Zhan$^{a}$}
\affiliation{$^{a}$State Key Laboratory of Magnetic Resonances and
Atomic and Molecular Physics, Wuhan Institute of Physics and
Mathematics, Chinese Academy of Sciences, Wuhan 430071, China}
\affiliation{$^{b}$School of Mathematics and Physics, China
University of Geosciences, Wuhan 430074, P.R. China}
\begin{abstract}
We consider a recent scheme of gravitational wave detection using atomic
interferometers as inertial sensors, and reinvestigate its configuration using
the concept of sensitivity functions. We show that such configuration can
suppress noise without influencing the gravitational wave signal. But the
suppression is insufficient for the direct observation of gravitational wave
signals, so we analyse the behaviour of the different noises influencing the
detection scheme. As a novel method, we study the relations between the
measurement sensitivity and the distance between two interferometers, and find
that the results derived from vibration noise and laser frequency noise are in
stark contrast to that derived from the shot noise, which is significant for
the configuration design of gravitational wave detectors using atomic interferometers.

\end{abstract}
\maketitle

\section{Introduction}

The detection of gravitational waves is a central topic for General
Relativity, and is also one of the most challenging efforts in experimental
physics at present \cite{gvlb13}. Light interferometers have been used as the
main tools to search for gravitational waves \cite{ba09,ta12}. Recently, due
to the highly developed control and manipulation techniques for atoms, a new
gravitational wave detection scheme, atomic gravitational wave interferometric
sensor (AGIS) \cite{dgr08,dgr09,plb11,dgr11,hjk11,plb12}, had been put
forward, which works with a similar mechanism to Laser Interferometer
Gravitational-Wave Observatory (LIGO) but replacing the macroscopic mirrors
with freely falling atoms. It was noted that in the original papers
\cite{dgr08,dgr09}, the principle of AGIS was applied in several different
implementations and configurations, such as a
Laser-Interferometer-Space-Antenna (LISA)-like three satellite configuration
for comparing it with LISA. In the paper, we will only consider the single-arm
configuration similar to that described in the Figure 4 of Ref. \cite{dgr08}.
In particular, when we mention the AGIS configuration in this paper, it only
refers to the single-arm configuration which will also be described in the
next section.

The advantage of the single-arm configuration using atomic interferometers as
inertial sensors \cite{dgr08,dgr09} is to suppress many kinds of background
noise significantly, but without reducing the gravitational wave signal. In
this type of gravitational wave detection schemes, some advances in detection
schemes have been made. Firstly, the proposal of the one-laser configuration
\cite{yt11} enables the use of time-delay interferometry to be extended to the
two atomic interferometers and thus cancels laser frequency noise but without
influencing the gravitational wave signal. Then, an intelligent proposal
\cite{ghkr13} including the large momentum transfer in this process had also
been put forward, with the suppression of laser frequency noise being derived
from the same principle \cite{yt11} that a single laser is used to manipulate
the two interferometers at different times. Although the one-laser
configuration has the potential to reduce the requirement of laser fractional
frequency stability, but some more challenges might prevent the application of
this kind of interferometers in the near future \cite{plb14}. On the other
hand, due to the development of atomic interferometers, the AGIS scheme is
more interesting, which is also discussed in a recent review on gravitational
wave detection \cite{rxa14}. Therefore, in this paper we will revisit the
detection scheme of AGIS and study the relationship between sensitivity and
the distance separating two atomic interferometers.

Besides the advantage of atomic interferometer that the freely falling atoms
can avoid the influence of vibration to a large extent, the initial aim of
AGIS is to allow the observation of low frequency sources in the
band$10^{-3}-10$ Hz, which fill up the gap between the detections from LIGO
and LISA and have many exciting astrophysical and cosmological sources
\cite{dgr08,dgr09}. So the detection of gravitational waves using atomic
interferometers is still a promising direction, although the detection
sensitivity of this kind of interferometers has to be improved further.
Remarkably, an experiment from our lab have improved the sensitivity for test
of equivalence principle using atomic interferometer \cite{zwz15}, and we have
also been trying to study the detection of gravitational waves using atomic
interferometers. In this paper, we will focus on the AGIS-like scheme and
study the influence of separating distance between atomic sensors for
gravitational wave detection using some atual experimental data for the
analysis of different kinds of noise.

Where gravitational wave detection was carried out using the AGIS, the signal
about gravitational waves is usually imprinted on the total phase difference
between the two spatially separated atomic interferometers, and the leading
term of the phase difference is proportional to the distance between the two
interferometers \cite{dgr08,dgr09}. Therefore, a larger distance between the
two interferometers is usually considered to be better for improving the
sensitivity of gravitational wave detection. However, that conclusion was made
after considering the effects of shot noise only. In this paper, we will
present a different and novel phenomenon in the change of sensitivity when the
distance between two interferometers varies, in the context where vibration
noise and laser frequency noise are considered. It is stressed that the
purpose of the paper is to study the relationship of sensitivity with the
distance between two interferometers, which will be significant for the design
of such the gravitational wave detector configurations as discussed later, and
may even be significant for the design of the atom interferometer-based
gravity gradiometers \cite{shk98}. In particular, it is also noted that the
distance between two interferometers in the single-arm configuration also
plays an important role for the comparison of atom interferometers and light
interferometers \cite{bt12}.

The structure of the paper is as follows. In the second section, we will
introduce the AGIS-like configuration that is used in this paper, and
interpretate the different ways entering the interferometers for gravitational
wave signal and different kinds of noise. Then we study the sensitivity
constrained by different kinds of noise in the third section. In the fourth
section we present the influence of the separating distance between two
interferometers on different kinds of noise. Finally, we give a conclusion in
the fifth section.

\section{Configuration for gravitational wave detection}

Our discussion is based on a configuration enclosed by the dashed box
presented in Fig.1, which is a schematic diagram similar to the AGIS scheme.
The mirror is used here since the AGIS uses two counterpropagating laser
pulses to manipulate the atomic interferometers. Thus the mirror will transfer
vibration noise to the passive laser beams which cannot be cancelled by a
common manipulation to the two atomic interferometers. In particular, a
similar mechanism to AGIS can be realised as such: at the time $t_{1}$ a pulse
with wavevector $k_{1}$ is emitted by the laser, and is reflected by the
mirror at the right side after its frequency is shifted through a frequency
shifter, which makes the effective wavevector $k=$ $k_{1}^{^{\prime}}-k_{1}$
match with the atoms going through the interferometers, where $k_{1}%
^{^{\prime}}$ is related to the pulse after the frequency is shifted. The
reflected pulse and the second pulse emitted at the time $t_{2}$ together
manipulate the right interferometer via a Raman process. A similar process is
implemented for the left interferometer, but with the third pulse emitted at
the time $t_{3}$ and the same reflected pulse. The times $t_{1}$, $t_{2}$, and
$t_{3}$ must be arranged carefully in order to make the process finished
exactly. Thus the first operation is finished, which is similar to the
beam-splitter operation in each interferometer. Then the second and the third
operations that similar to mirror and beam-splitter operations respectively
for every interferometer, are made with a time interval $T$ between two
successive operations.

\begin{figure}[ptb]
\centering
\includegraphics[width=3.25in]{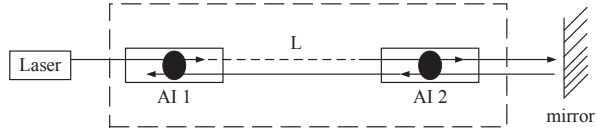}\caption{Schematic diagram of a
gravitational wave detector using atomic interferometers as local inertial
sensors}%
\end{figure}

Now, we will describe the configuration in Fig.1 using the sensitivity
function, and explain why some kinds of noise such as vibrational noise will
be suppressed, but without affecting the signal arising from a gravitational
wave. The sensitivity function is suggested firstly by Dick \cite{gjd87,sac98}%
, and then investigated in detail by others \cite{ccl08,gms08,bgb13,bzz13}
with a time-domain atomic interferometer \cite{kc91}, which quantifies the
influence of a relative laser phase shift $\delta\phi$ occurring at a time $t$
during the interferometer sequences on the transition probability $\delta
P\left(  \delta\phi,t\right)  $; it is then defined in Ref. \cite{gjd87,sac98}
as%
\begin{equation}
g\left(  t\right)  =2\ast\underset{\delta\phi\rightarrow0}{\lim}\frac{\delta
P\left(  \delta\phi,t\right)  }{\delta\phi}.\label{sf}%
\end{equation}
If the time origin is chosen at the middle of the second Raman pulse, the
sensitivity function $g\left(  t\right)  $ is an odd function. For the three
pulses $\frac{\pi}{2}-\pi-\frac{\pi}{2}$ with durations respectively
$\tau-2\tau-\tau$, we choose the initial time $t_{i}=-T$ and the final time
$t_{f}=T$ to obtain the expression of the sensitivity function of the left
interferometer as \cite{ccl08,gms08}
\begin{equation}
g_{1}(t)=\left\{
\begin{array}
[c]{c}%
\sin(\Omega(T+t))\ \ \ ,-T\leqslant t<-T+\tau\\
\ 1\ \ \ \ \ \ \ \ \ \ \ \ \ ,-T\ +\tau\leqslant t<-\tau\\
-\sin\Omega t\ \ \ \ \ \ \ \ ,-\tau\leqslant t<\tau\ \ \ \ \ \ \\
-1\ \ \ \ \ \ \ \ \ \ \ \ ,\tau\leqslant t<T-\tau\ \ \ \\
-\sin\Omega(T-t)\ \ \ \ ,T-\tau\leqslant t\leqslant T\ \ \
\end{array}
\right.  \label{1s}%
\end{equation}
where $\Omega$ is the effective Rabi frequency, $T$ is the interrogation time
between the two sequential pulses, and $g_{1}(t)=0$ for $\left\vert
t\right\vert >T$ due to the phase jump occurring outside the interferometer.

According to the configuration in Fig.1, the interferometer on the right is
operated with the same reflected laser pulse as the one on the left, and so
the interference time is earlier on the time axis than that for the left one,
so its sensitivity function is expressed as,%
\begin{equation}
g_{2}(t)=\left\{
\begin{array}
[c]{c}%
\sin(\Omega(T+t+L))\ \ \ ,-T-L\leqslant t<-T+\tau-L\\
\ 1\ \ \ \ \ \ \ \ \ \ \ \ \ ,-T+\tau-L\leqslant t<-\tau-L\\
-\sin\Omega\left(  t+L\right)  \ \ \ \ \ \ \ \ ,-\tau-L\leqslant
t<\tau-L\ \ \ \ \ \ \\
-1\ \ \ \ \ \ \ \ \ \ \ \ ,\tau-L\leqslant t<T-\tau-L\ \ \ \\
-\sin\Omega(T-t+L)\ \ \ \ ,T-\tau-L\leqslant t\leqslant T-L\ \ \
\end{array}
\right.  \label{2s}%
\end{equation}
where we have taken the speed of light $c=1$. Note that the function
$g_{2}(t)$ is not odd in this situation. Then, the sensitivity function of
differential measurement configurations at a specific time $t$ can be found as
$g_{W}(t)=g_{2}(t)-g_{1}(t)$. After Fourier transformation, the transfer
function is obtained as,
\begin{align}
H(\omega) &  =\omega\int e^{-i\omega t}g_{W}(t)dt\nonumber\\
&  =\frac{4\Omega\sin\frac{\omega T}{2}\left(  \omega\cos\frac{\omega T}%
{2}+\Omega\sin\frac{\omega\left(  T-2\tau\right)  }{2}\right)  }{\omega
^{2}-\Omega^{2}}\left[  \sin\omega L+i\left(  1-\cos\omega L\right)  \right]
\label{tf}%
\end{align}
where the first term is the transfer function of a single atomic
interferometer \cite{ccl08,gms08} and the second term is related to the time
delay of the light pulse. Note that the sensitivity function is calculated at
the same time for the two interferometers, which means the pulses used to
manipulate the two interferometers is independent. Thus the laser intensity
noise would make the Rabi frequencies different for the two interferometers,
meaning that $\Omega_{1}\neq\Omega_{2}$. According to the present laser power
stabilisation technology \cite{kwd11} and previous noise analyses
\cite{gms08,pcc01} for atomic interferometers, the laser intensity noise is
sufficiently small that it can be neglected. Therefore, in this paper we
ignore the influence of laser intensity noise on the expression of the
sensitivity function.

As a consistency check, we can recreate Figure 6 of Ref. \cite{dgr08} with the
same parameters by using our transfer function in Eq. (\ref{tf}). On the other
hand, the sensitivity functions in Eqs. (\ref{1s}) and (\ref{2s}) do not
indicate any differences between manipulations using two Raman pulses and
using a single pulse \cite{yt11,ghkr13}, so we expect the differential
measurement for the one-laser configuration to give a zero result, achieved by
considering a common laser pulse to manipulate the two interferometers; in
other words, the configurational sensitivity function should be zero for such
a measurement. Assume that $\varphi$ is a phase change caused by vibration
noise, and we have the differential expression for the same laser beams,
$\Delta\varphi=\int\left(  g_{1}(t)\frac{d\varphi\left(  t\right)  }{dt}%
-g_{2}(t)\frac{d\varphi\left(  t-L\right)  }{dt}\right)  dt=\int\left(
g_{1}(t+L)-g_{2}(t)\right)  \frac{d\varphi\left(  t\right)  }{dt}dt=0$, as
expected in Ref. \cite{yt11}. However, for the configuration described in
Fig.1, if such cancellation is considered for the same reflected laser pulses,
there will not be the same cancellation for the incoming ones, and vice versa,
as discussed for AGIS in Ref. \cite{dgr08}. The model constructed here can
illustrate this effect if we choose the time delay properly: the transfer
function in Eq. (\ref{tf}) is chosen for the configuration in Fig.1 or AGIS,
and $H^{^{\prime}}(\omega)=\omega\int\left(  g_{1}(t+L)-g_{2}(t)\right)
e^{-i\omega t}dt$ is chosen for the one-laser configuration, but note that
$H^{^{\prime}}(\omega)$ would not be zero if the relative motion of the two
interferometers is considered \cite{ghkr13}.

In order to present the advantage of this configuration, we take the vibration
noise (see Ref. \cite{hsa13} for a detailed analysis of vibration noise) as an
example and observe how much the noise is suppressed in the differential
measurement. Assuming that the noise is brought into the detection device
through coupling to the mirror or the laser platform, the influence of
vibration noise on the total phase can be estimated by the variance in phase
fluctuation \cite{ccl08,gms08},%
\begin{equation}
\sigma_{\varphi}^{2}=\int_{0}^{\infty}\left\vert H_{\varphi}(\omega
)\right\vert ^{2}S_{\varphi}(\omega)d\omega=\frac{k^{2}}{\omega^{4}}\int
_{0}^{\infty}\left\vert H_{\varphi}(\omega)\right\vert ^{2}S_{a}%
(\omega)d\omega\label{ns}%
\end{equation}
where $k$ is the effective laser-field wavevector. Fig.2 is a vibration
spectrum measured in our lab, without any isolation system added in the
measurement process. With this vibration spectrum, we can estimate the
influence of vibration noise on the final phase difference. For example, with
the parameters $T=1.4$ s, $\tau=4\times10^{-5}$ s, $k=1\times10^{7}$ m$^{-1}$,
$L=1000$ m, the differential value of the final phase shift is approximately
$1.2\times10^{-5}$ rad, which is significantly suppressed compared with the
result obtained from a single interferometer, that is approximately $2.94$ rad.

\begin{figure}[ptb]
\centering
\includegraphics[width=3.25in]{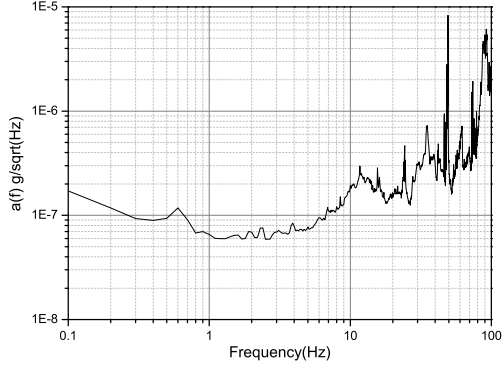}\caption{The spectrum of the vibration
measured in our lab using a seismometer, without any isolation system included
in the measurement process. a(f) is the acceleration noise power spectral
density.}%
\end{figure}

However, the signal from gravitational waves is not obtained directly from the
result by the differential measurement. The signal can simply be obtained
through two steps: firstly the gravitational wave induces phase fluctuations
of light propagating along the baseline $L$ of our experimental setup, and
secondly, the phase fluctuations are transferred into the atoms through our
constructed configuration. Assuming that the gravitational wave is propagating
along the direction perpendicular to the baseline $L$, and choosing
coordinates carefully, for example, we choose the propagating direction of
gravitational wave to be along the Z direction, and the baseline is along the
X direction (see Ref. \cite{ew75} for the details). Thus the influence on the
laser phase of gravitational waves travelling through the proposed detector is
expressed as%
\begin{equation}
\frac{1}{k}\frac{d\phi_{G}\left(  t\right)  }{dt}=\frac{1}{2}\left[
h_{+}\left(  t-L\right)  -h_{+}\left(  t\right)  \right]  \label{lgp}%
\end{equation}
where the gravitational wave presents only the \textquotedblleft%
+\textquotedblright\ polarisation in the transverse-traceless gauge. Without
loss of generality, we take $h_{+}\left(  t\right)  =h\sin\left(
\upsilon\left(  t+\frac{L}{2}\right)  +\phi_{0}\right)  $ where $h$ is the
amplitude of the gravitational waves, and $\phi_{0}$ is an arbitrary initial
phase. Then, using the sensitivity function for the right interferometer, we
obtain the phase shift from the signal of gravitational wave as%
\begin{align}
\Delta\phi_{G} &  =\int g_{2}(t)\frac{d\phi_{G}\left(  t\right)  }%
{dt}dt\nonumber\\
&  \simeq2\frac{hk}{\nu}\sin^{2}\left(  \frac{\upsilon T}{2}\right)
\sin\left(  \frac{\upsilon L}{2}\right)  \sin\left(  \frac{\upsilon L}%
{2}+\upsilon T+\phi_{0}\right)  \label{sp}%
\end{align}
which is the same as the result obtained in Ref. \cite{dgr08} since the
effective wavevector $k=k_{2}-k_{1}\simeq2k_{2}$.

Thus from Eqs. (\ref{ns}) and (\ref{sp}), it is easy to see that a class of
common noises is significantly suppressed, but the signal is not reduced by
the differential measurement since it is detected only by the right
interferometer, which included an implicit assumption that the distance
between the laser and the left interferometer is much less than $L$.

\section{Sensitivity constrained by noise}

In the last section, we use the sensitivity function to analyse the
configuration described in Fig.1, and in this section we investigate how
different kinds of noise influence the sensitivity of such a configuration. In
particular, we compare the difference in sensitivities constrained by shot
noise with that arising from the phase noise which is caused by vibrations and
laser frequency instability. The vibration spectrum, as an example, is shown
in Fig.2 and the laser frequency spectrum is shown in Fig.3, both of which are
experimentally obtained in our lab. In what follows, we will calculate these
sensitivities and get their corresponding curves.

\begin{figure}[ptb]
\centering
\includegraphics[width=3.25in]{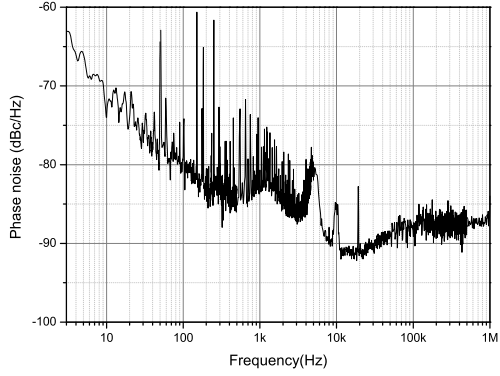}\caption{The spectrum of the laser
frequency in our lab}%
\end{figure}

In the AGIS scheme, the large-momentum-transfer (LMT) beam splitters that
consist of a $\frac{\pi}{2}$ pulse and $N$ pairs of $\pi$ pulses and the LMT
mirror that included $2N+1$ $\pi$ pulses in the sequence are used to
manipulate the atomic interferometers with a Bragg process \cite{mcc07,ckk11}.
Without affecting our purpose in this paper, we chose $N=1$. Therefore, the
sensitivity function of the left interferometer is expressed as
\begin{equation}
g_{1}(t)=\left\{
\begin{array}
[c]{c}%
\sin\Omega(t+T)\ \ \ \ \ \ \ \ \ \ \ \ \ \ -T\leqslant t<-T+\tau\\
\ \ \ \ \ \ \ \ \ 1\ \ \ \ \ \ \ \ \ \ \ \ \ \ \ \ \ \ \ -T+\tau\leqslant
t<-T+3\tau\\
\frac{1}{2}[3-\cos\Omega(T+t-3\tau)]\ \ \ \ \ \ -T+3\tau\leqslant t<-T+5\tau\\
\ \ \ \ \ \ 2\ \ \ \ \ \ \ \ \ \ \ \ \ \ \ \ \ \ \ -T+5\tau\leqslant
t<-5\tau\\
\frac{1}{2}[3+\cos\Omega(t+5\tau)]\ \ \ \ \ \ \ \ \ -5\tau\leqslant
t<-3\tau\ \ \ \ \ \ \ \\
\ 1\ \ \ \ \ \ \ \ \ \ \ \ \ \ \ \ \ \ \ -3\tau\leqslant t<-\tau\\
\ -\sin\Omega t\ \ \ \ \ \ \ \ \ \ \ \ \ \ \ \ -\tau\leqslant t<0\ \ \ \ \ \
\end{array}
\right.
\end{equation}
written for the time period of $t<0$ since we can obtain an odd function when
we choose the time origin to be at the middle of the middle $\pi$ pulse. From
the last section, we know that $g_{2}(t)=g_{1}(t+L)$ without considering the
relative motion of the two interferometers. Thus the differential sensitivity
function $g_{W}(t)=g_{2}(t)-g_{1}(t)$ and its transfer function is obtained
through the same formula $H(\omega)=\omega\int e^{-i\omega t}g_{W}(t)dt$. We
have also checked the transfer function by comparing it with Figure 6 of Ref.
\cite{dgr08}, generating the figure with the same parameters as those used in
the paper, and the nearly unchanged result shows that the introduction of LMT
does not influence the passband of the device which is truncated at the
frequency of $\frac{1}{L}$ and the Rabi frequency $\Omega$.

As shown in Ref. \cite{dgr08}, the noise will not be amplified, since all but
the beginning and end of each LMT pulse will be common to the two
interferometers, if the Rabi frequency $\Omega$ and the distance $L$ between
the two interferometers were chosen properly. That means that the manipulation
using LMT pulses will increase the measurement sensitivity in a way
proportional to $N$. In general, the sensitivity of gravitational wave
detection can be obtained by the equation,
\begin{equation}
SNR=\frac{\Delta\phi_{G}}{\sigma_{\varphi}}\label{snr}%
\end{equation}
which is the ratio of signal to noise, and the signal $\Delta\phi_{G}$ has
included the result of LMT. In Ref. \cite{dgr08}, the detection sensitivity
constrained by shot noise for some range of frequencies was better for a
larger distance $L$, but the frequency range was found to be dependent on the
distance $L$. In other words, when the distance between the two
interferometers varies, the corresponding optimal detection frequencies will
also vary, which is the requirement of the approximation taken in Ref.
\cite{dgr08}. In this paper, we do not take such approximation for the leading
term of the calculated result, as seen in the expression of Eq. (\ref{sp}).
Our result was presented in Fig.4, and a comparison of two separation
distances indicates that a larger separation distance could improve the
sensitivity, which is shown clearly in the next section. In particular,
according to our estimation for shot noise, the distance has to be as large as
$10^{8}$ m if the amplitude of gravitational wave is of the order of
$h\sim10^{-20}$.

\begin{figure}[ptb]
\centering
\includegraphics[width=3.25in]{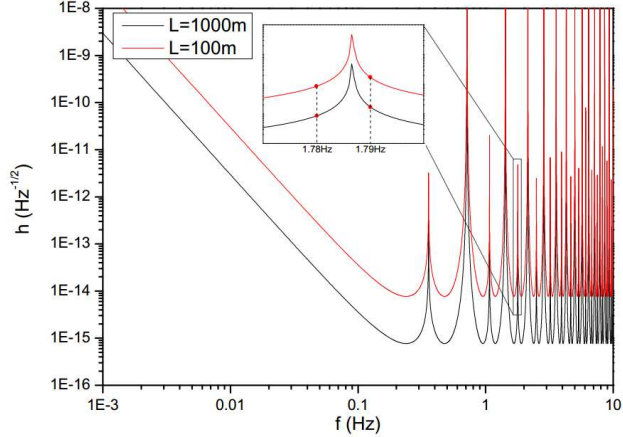}\caption{The sensitivity curve to
gravitational wave of frequency f, limited by shot noise. The different
parameters are labeled in the diagram. The common parameters that were used:
$\tau=4\times10^{-5}$s, $k=1\times10^{7}$m$^{-1}$. These are similar to the
proposal in the terrestrial experiment of Ref. \cite{dgr08}. }%
\end{figure}

The sensitivity curves related to the vibration noise and laser frequency
noise have not been investigated before, and here we also plot these values in
Fig.5. Surprisingly, the sensitivity is nearly unchanged when the distance
between the two interferometers is increased. This is consistent with the
analysis for the vibration noise in Ref. \cite{bt12}, where the ratio of
signal to vibration noise is independent on the distance between two
interferometers, under the approximation of $\omega L\ll1$. But for the
single-arm light interferometer, this ratio is proportional to $1/L^{2}$
\cite{bt12}, so it is necessary to investigate further the influence of the
distance between two interferometers.

\begin{figure}[ptb]
\centering
\includegraphics[width=5.25in]{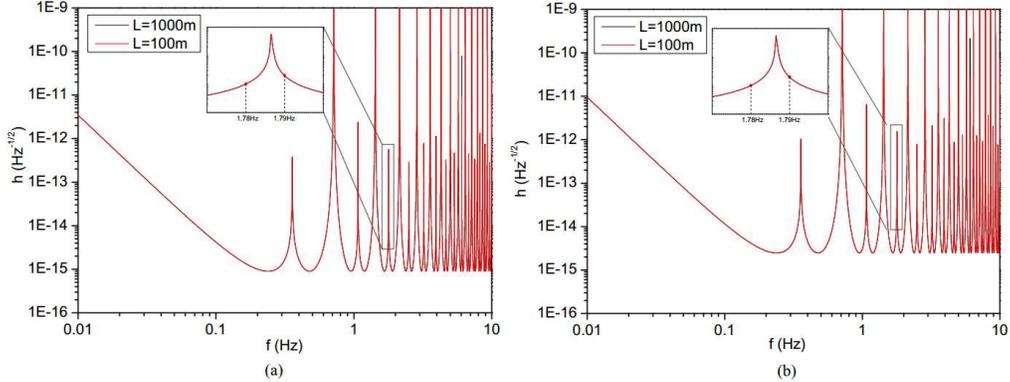}\caption{The sensitivity curve to
gravitational wave of frequency f, limited by vibration noise (a) and laser
frequency noise (b). The parameters are same as that in Fig.4. }%
\end{figure}

\section{Influence of the distance between two interferometers}

In gravitational wave detection, the relation of sensitivity with the
measurable frequencies plays an important role for the estimation of the
detection scheme. From the last section, we notice that sensitivity curves
almost look the same for the shot noise, vibration noise, and laser-frequency
noise, and an obvious difference is the change in sensitivity when the
distance between two interferometers changes. In this section, we study this
relation in detail without any approximation for the leading term, and present
the difference of influence on gravitational wave detection arising from
vibration noise or laser frequency noise from that arising from shot noise.

For the shot noise, the sensitivity will be better for a larger distance $L$,
which is confirmed by studying the relation of measurable amplitude of
gravitational wave with the distance $L$, as in Fig.6a. It was noticed that
the sensitivity curves is different for the two chosen frequencies $f=1.78$ Hz
and $f=1.79$ Hz (they are labeled in Fig.4 with two different red points) at
any given distance $L$, but the changing trends are the same for two curves.
Actually, for the shot noise, the changing trend is the same for any given
frequency, although the sensitivity is different at some given distance $L$.
Moreover, the sensitivity presented here seemed worse than what would be
required for gravitational wave detection, because the parameter $k$ was
chosen to be a value smaller than that in Ref. \cite{dgr08}. It is easy to see
that if we take the parameter $k$ with the same value with that in Ref.
\cite{dgr08}, the sensitivity will nearly reach $10^{-17}$ at the distance
$L=1000$m. In particular, $k$ is considered as a free parameter due to LMT,
and doesn't have to be directly tied to the choice of atom and transitions.
Here the important thing is to understand the relation of sensitivity with the
separation distance between two interferometers, and in what follows we give a
different form of this relation when the vibration noise and laser-frequency
noise are considered.

\begin{figure}[ptb]
\centering
\includegraphics[width=6.25in]{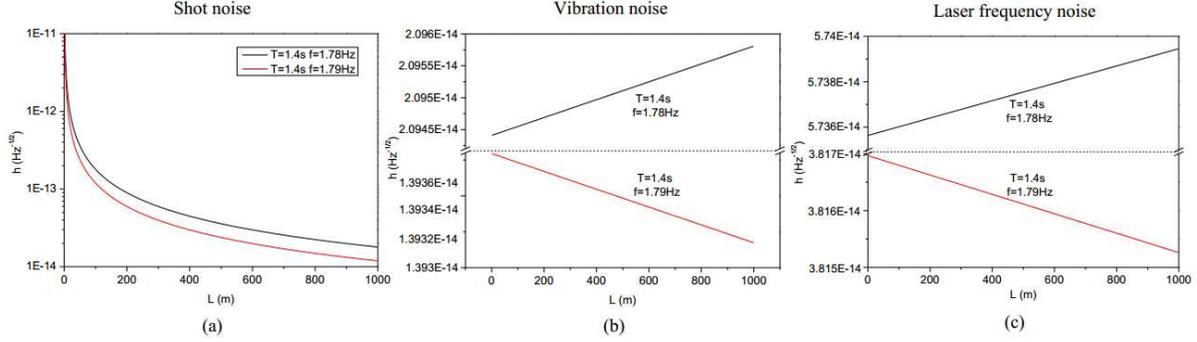}\caption{The relation of sensitivity
with the distance $L$ between two interferometers, limited by shot noise (a),
vibration noise (b), and laser frequency noise (c). The common parameters are
still used: $\tau=4\times10^{-5}$s, $k=1\times10^{7}$m$^{-1}$. The different
frequency parameters are labeled in the diagram, which are corresponding to
the points labeled in Fig.4 and Fig.5. }%
\end{figure}

As stated in the last section, the vibration noise arises from the laser
platform or the mirror and coupled into the final results with different
proportions for different frequencies, as seen from the transfer function
(\ref{tf}). Then we take $\sigma_{\phi}$ in Eq. (\ref{snr}) as the
perturbation of vibration noise to determine the sensitivity. From Fig.6b, it
can be seen that the changing trend of the sensitivity with the separation
distance between two interferometers is different for the two chosen
frequencies $f=1.78$ Hz and $f=1.79$ Hz. Actually, for every frequency in
sensitivity curves of Fig.5, the changing trend of the sensitivity with the
separation distance have the same lineshape with either that of frequencies
$f=1.78$ Hz or that of frequencies $f=1.79$ Hz. This is subtly different from
the result obtained in Ref. \cite{bt12} but is not in conflict with each
other, since the change in Fig.6b is small. However, the result presented in
Fig.6b shows that a larger distance does not improve the sensitivity, which is
related to noise. Our result is also applicable to other common noises such as
laser phase noise etc., which is easy to understand since the influence of
vibration on the final result is similar to the influence from phase noises
\cite{mjv06}. It was noted that the sensitivity is far from what is required
for gravitational wave detection, since the vibration spectrum in Fig.2 is
measured without using any vibration-isolation system in the measurement
process, and we have given a rough estimate for its amplitude in the second
section. Moreover, the different values of sensitivity for two different
frequencies can be explained with reference to the case of shot noise.
However, the distinctly different behaviours in relation to separation
distance between Fig. 6b and Fig. 6a shows that vibration noise influences the
final measurement result with a quite different way from shot noise.

From Fig.6c, we see that the result derived from the laser frequency noise,
which also contributes to the laser phase noise, is almost the same as that
from the vibration noise. Laser frequency noise is a dominant background noise
for gravitational wave detection using light interferometers, and it could be
suppressed by applying single laser pulse to operate the two atomic
interferometers simultaneously \cite{yt11,ghkr13} although there is still some
technological difficulties with this kind of measuring scheme \cite{plb14}.
For our scheme under consideration in Fig.1, although the reflected laser
pulse is the same for the two interferometers, the incoming ones are
different, which is main source of laser frequency noise. Here, to compare the
behaviour of the laser frequency noise influencing the detection configuration
shown in Fig.1 with other types noise, we use the laser frequency spectrum in
Fig.3, like the vibration spectrum, without using any methods to suppress the
laser frequency noise. It is noted that the sensitivity varies slightly with
the distance between two interferometers, although the phase change caused by
laser frequency noise will increase linearly when the distance is lengthened,
which is approximately estimated by an approximate estimation as $\delta
k\cdot L$ where $\delta k$ is due to the average deviation of the laser
frequency over the finite time length of the pulse. This amplification will
nearly cancel the increase in the signal from gravitational waves, and thus
leads to the resulting slight change in sensitivity.

Actually the most significant estimation for practical observations lies in
the whole average noise, that is $\sigma_{\Delta\varphi}=\sqrt{\sigma
_{\phi_{i}}^{2}+\sigma_{\phi_{j}}^{2}+\sigma_{0}^{2}}$ where $\sigma_{0}^{2}$
is from shot noise, $\sigma_{\phi_{i}}^{2}$ are from known noise sources such
as phase noise etc., and $\sigma_{\phi_{j}}^{2}$ are from potential systematic
effects such as some environmental effects. According to the level of noise,
given here by shot noise and vibration noise, the sensitivity limited by their
average noise may still be improved by a large distance $L$, but this
improvement is dependent on which type of noise is dominant in the
observation, for example, in the 1999 experiment to measure gravitational
acceleration by dropping atoms, the phase noise is about a hundred times
larger than atomic shot noise \cite{pcc99,zcz13}. Thus, if phase noise is
dominant in the future gravitational wave observation, whether the separation
distance between two interferometers should be made larger has to be
considered carefully. That is to say, a more economic selection for the
distance $L$ may be possible, dependent on the actual level of noise.

In particular, we pointed out that although our analysis was carried out using
the parameters $T=1.4$ s, $\tau=4\times10^{-5}$ s, $k=1\times10^{7}$ m$^{-1}$
which will determine the passband of the single-arm scheme of gravitational
wave detectors, the conclusion will not change if we take other physically
allowed values for these parameters. Moreover, $L$ is also an factor that can
determine the passband, but from Fig. 6b and 6c, we find that our conclusions
are not influenced by the positions of the passband. On the other hand,
although our analysis is for the ground-based single-arm scheme of
gravitational wave detection, it is still meaningful and significant for the
similar space-based scheme, and in particular, for the space-based situation,
the vibration noise is stronger \cite{ac04,zfc12} and the laser frequency
noise might be more difficult to be overcome. If the spectra for measuring
vibration or laser frequency instability for a space-based situation was used
to do the same calculation with that for ground-based ones, the corresponding
results can also be obtained, and in our view it is highly possible to arrive
at the same conclusions as presented in Fig.6.

\section{Conclusion}

In this paper, we have described the sensitivity function for single-arm
gravitational wave detector based on atomic interferometers, in which we
explain the way that a range of noise from different sources, such as
vibration noise and laser frequency noise, can enter the detector, but in a
different way from that the signal arising from gravitational waves enters the
detector. We have also studied the relation of the measurement sensitivity
with the distance between two interferometers, and showed that the sensitivity
limited by these common sources of noise will nearly be unchanged when the
distance is greatly increased, which is significantly different from the case
for shot noise. Then a next significant thing is to investigate the
sensitivity which is limited by other kinds of noise, in particular, by the
Newtonian gravity background which might influence the sensitivity with a
different way, as the gravity background fluctuates in value depending on
position and time.

\begin{acknowledgments}
We thank Dr. J. M. Hogan for his helpful and detailed explanation for his
papers. Financial support from NSFC under Grant Nos. 11374330, 11104324 and
11227803, and NBRPC under Grant No.2010CB832805 is gratefully acknowledged.
\end{acknowledgments}

\end{document}